# Statistical Distortion: Consequences of Data Cleaning

Tamraparni Dasu
AT&T Labs Research
180 Park Avenue
Florham Park, NJ 07932
tamr@research.att.com

Ji Meng Loh
AT&T Labs Research
180 Park Avenue
Florham Park, NJ 07932
loh@research.att.com

## ABSTRACT

We introduce the notion of *statistical distortion* as an essential metric for measuring the effectiveness of data cleaning strategies. We use this metric to propose a widely applicable yet scalable experimental framework for evaluating data cleaning strategies along three dimensions: glitch improvement, statistical distortion and cost-related criteria. Existing metrics focus on glitch improvement and cost, but not on the statistical impact of data cleaning strategies. We illustrate our framework on real world data, with a comprehensive suite of experiments and analyses.

## 1. INTRODUCTION

Measuring effectiveness of data cleaning strategies is almost as difficult as devising cleaning strategies. Like data quality, cleaning metrics are highly context dependent, and are a function of the domain, application, and user. To the best of our knowledge, statistical distortion caused by data cleaning strategies has not been studied in any systematic fashion, and is not explicitly considered a data quality metric. Existing approaches focus on data quality metrics that measure the effectiveness of cleaning strategies in terms of constraint satisfaction, glitch improvement and cost. However, data cleaning often has serious impact on statistical properties of the underlying data distribution. If data are over-treated, they no longer represent the real world process that generates the data. Therefore, cleaner data do not necessarily imply more useful or usable data.

Any change to the data could result in a change in the distributional properties of the underlying data. While there are situations where changing the distribution is not of concern, in many cases changing the underlying distribution has an impact on decision making. For instance, a change in the distribution could result in an increase in false positives or false negatives because the distribution is now out of sync with the thresholds already in place, such as those used for alerting loads on the network.

There is considerable research in the database community on varied aspects of data quality, data repair and data cleaning. Recent work includes the use of machine learning for guiding database repair [14]; inferring and imputing missing values in databases [10] ; resolving inconsistencies using functional dependencies [6]; and for data fusion [8].

From an enterprise perspective, data and information quality assessment has been an active area of research as well. The paper [12] describes subjective and objective measures for assessing the quality of a corporation's data. An overview of well known data quality techniques and their comparative uses and benefits is provided in [2]. A DIMACS/CICCADA workshop [1] on data quality metrics featured a mix of speakers from database and statistics communities. It covered a wide array of topics from schema mapping, graphs, detailed case studies and data privacy. Probability and uncertainty have been used in identifying and developing cleaning strategies for a wide range of tasks such as missing value imputation, inconsistency and duplicate resolution, and for the management of uncertain data. However, to the best of our knowledge, there is no known data quality metric that measures the impact of cleaning strategies on the statistical properties of underlying data.

### 1.1 Data Distributions Matter

As a simple though egregious example, consider the case where there are two distributions with the same standard deviation $\sigma$. A blind data cleaning strategy is used to clean the data: apply 3-$\sigma$ limits to identify outliers, where any data point outside the limits is considered an outlier; "repair" the outliers by setting them to the closest acceptable value, a process know as *Winsorization* in statistics.

In Figure 1, the schematic histogram (top left) represents the distribution assumed for developing the outlier rule and repair strategy. Outliers are shown in dark brown at the two extreme tails of the distribution. The schematic histogram at top right represents the actual data. Suspicious data, in the form of potential density-based outliers in a low density region, are shown in orange. When we apply the outlier rule to the actual data (bottom left), the two bars on either extreme are identified as outlying data. On repairing the data using Winsorization, we end up with a data distribution shown in the bottom right histogram.

There are several concerns with this approach. First, legitimate "clean" data were changed, as shown schematically by the dark blue bars in the bottom right histogram in Figure 1. What is worse, legitimate values were moved to a potentially outlying region (above the orange bar). Therefore this data cleaning strategy that was developed independently of the data distribution of the dirty data set, resulted in a fundamental change in the data, moving it distributionally far from the original. Such a radical change is unacceptable in many real world applications because it can lead to incorrect decisions. For example, in the cleaned data, a significant portion of the data lie in a low-likelihood region of the original, adding to the number of existing outliers. If the low-likelihood region corresponds to forbidden values (the orange values might correspond to data entry errors), our blind data repair has actually introduced new glitches and made the data dirtier.

[1]http://dimacs.rutgers.edu/Workshops/Metrics/

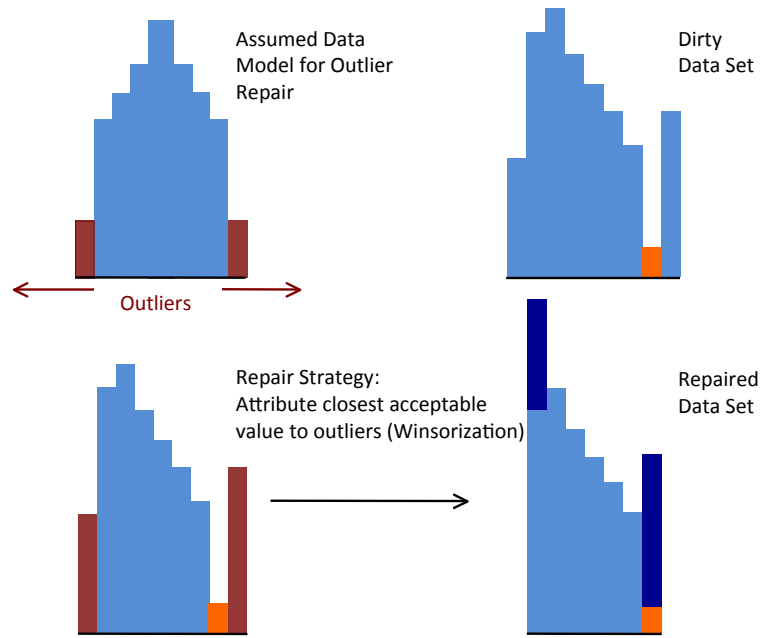

Figure 1: Outliers (dark brown) are identified using 3-$\sigma$ limits assuming a symmetric distribution; They are repaired by attributing the closest acceptable (non-outlying) value called Winsorization. This approach results in errors of commission (legitimate values are changed, dark blue) and omission (outliers are ignored, orange). The data repairs change the fundamental nature of the data distribution because they were formulated ignoring the statistical properties of the data set.

Our goal is to repair the data to an ideal level of cleanliness while being statistically close to the original, in order to stay faithful to the real life process that generated the data. Hence, we measure distortion against the original, but calibrate cleanliness with respect to the ideal. In order to achieve our goal, we introduce a three-dimensional data quality metric based on glitch improvement, statistical distortion and cost. The metric is the foundation of an experimental framework for analyzing, evaluating and comparing candidate cleaning strategies.

## 2. OUR APPROACH

We detail below our contributions, and provide an intuitive outline of our approach and methodology. We defer a rigorous formulation to Section 3.

### 2.1 Our Contribution

We make several original contributions: (1) we propose *statistical distortion*, a novel measure for quantifying the statistical impact of data cleaning strategies on data. It measures the fidelity of cleaned data to original data which we consider a proxy for the real world phenomenon that generates the data; (2) we develop a general, flexible experimental framework for analyzing and evaluating cleaning strategies based on a three dimensional data quality metric consisting of statistical distortion, glitch improvement and cost. The framework allows the user to control the scale of experiments with a view to cost and accuracy; and (3) our framework helps the user identify viable data cleaning strategies, and choose the most suitable from among them.

In general, the problem of data cleaning can be specified as follows: Given a data set $\mathcal{D}$, we want to apply cleaning strategies such that the resulting clean data set $\mathcal{D}_C$ is as clean as an ideal data set $\mathcal{D}_I$. The ideal data set could be either pre-specified in terms of a tolerance bound on glitches (e.g. "no missing values, less than 5% outliers"), or in terms of a gold-standard it must resemble, e.g., historically gathered data. Usually, strategies are chosen based on cost and glitch improvement. However, cleaning strategies introduce distributional changes that distort the statistical properties of the original data set $\mathcal{D}$, as discussed in our hypothetical example of Section 1.1. We propose a novel data quality metric called *statistical distortion* that takes this into account.

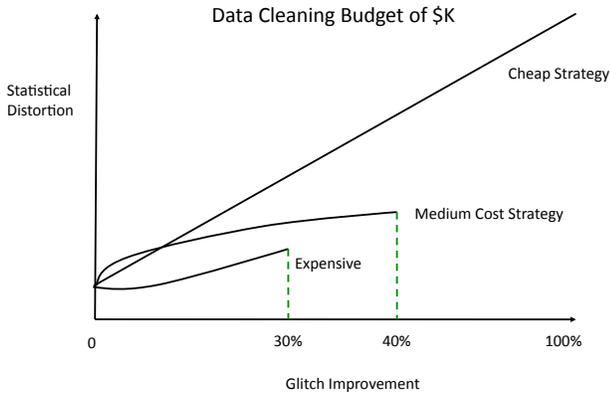

**Figure 2:** Given a data cleaning budget of $K, a user can choose between strategies that provide a trade-off between glitch improvement and statistical distortion.

The framework for our approach consists of three components: (1) an *experimental setup* for evaluating cleaning strategies; (2) glitch scoring for assessing data quality; and (3) measuring *statistical distortion* to identify viable strategies. The user then chooses a viable strategy that best suits the project, subject to cost and efficiency considerations.

As an example consider Figure 2. The user has a fixed budget $K. For simplicity, assume that the only types of glitches are missing values. Suppose they are imputed using a fixed constant, like the mean. This is an inexpensive strategy, and results in a 100% glitch improvement as all the glitches can be removed. But the data set is now distorted, since there is a spike in density at the mean of the distribution. Suppose, we use a medium cost strategy that enables us to simulate the distribution, but we have computing resources to cover only 40% of the glitches. The distortion in the data is much lower. Finally, suppose we re-take the measurements on the missing data and obtain exact values. This is even more expensive and can clean only 30% of the glitches, but the statistical distortion is lower. The "best strategy" depends on the user's tolerance to missing values: a user who is required by corporate mandate to have no missing values, will choose the cheap strategy, while another user who wishes to capture the underlying distribution will choose the expensive strategy.

### 2.1.1 Experimental Framework

Our experimental framework consists of repeated evaluations of strategies on small samples of data. We generate pairs of dirty and clean data sets by sampling with replacement from the dirty data set $\mathcal{D}$ and the ideal data set $\mathcal{D}_I$, to create the test pair $\{\mathcal{D}^i, \mathcal{D}_I^i\}$, $i = 1, \ldots R$. Each pair is called a *replication*, with $B$ records in each of the data sets in the test pair. The number of replications $R$ need not be very large. Replications give us multiple measurements on a given strategy $S$, and any value of $R$ more than 30 is sufficient to capture the behavior of $S$. The sample size $B$ (number of data records in each of the data sets in a test pair) is a function of cost, but need not be very large. The type of sampling can be geared to a user's specific needs by differential weighting of subsets of data to be sampled. Our focus in this paper is on measuring statistical distortion, and not on evaluating sampling strategies for large data bases. Please see [11] for random sampling from databases, [4] for bottom $k$-sketches sampling scheme, and [5] for the use of priority sampling for computing subset sums.

### 2.1.2 Ideal Data Set

There are two possible scenarios for identifying $\mathcal{D}_I$, the ideal data set. In the first one, there is a known ideal distribution or data set, denoted by $\mathcal{D}_I$. This could be a theoretically specified distribution ("multivariate Gaussian"), or a data set constituted from historically observed good data. In the second situation where there is no known ideal distribution or data set, we identify parts of the dirty data set $\mathcal{D}$ that meet the clean requirements ("no missing attributes, no more than 5% outliers") and treat these as the ideal data set $\mathcal{D}_I$, and generate $\mathcal{D}_I^i$ from this $\mathcal{D}_I$. To understand this, consider the data set as being sorted in terms of cleanliness. We would like the "dirty" part of the data to achieve the cleanliness of the "clean" section, without losing its distributional properties so as to control bias and statistical distortion. The user decides where to draw the line between clean and dirty, depending on the resources available for cleaning. For instance, in many network anomaly detection applications, duplicates and missing values are tolerated to a great extent if the sample of clean data is sufficiently large.

We apply a cleaning strategy $\mathcal{C}$ to $\mathcal{D}^i$ and obtain the *treated* or *cleaned* data set $\mathcal{D}_C^i$. We can now evaluate the cleaning strategy $\mathcal{C}$ using a glitch index to measure the improvement in data quality, and statistical distance between multidimensional distributions to measure statistical distortion.

### 2.1.3 Glitch Index

Let a data set $\mathcal{D}$ have $n$ records with $v$ attributes. In addition, let there be $m$ types of glitches that can be associated with a record, or an individual attribute in a record.

Within each cell of the $n \times v$ data matrix, we can denote the presence or absence of a glitch type by 1 or 0 respectively. Therefore, with every $d_{ij} \in \mathcal{D}$, we can associate a glitch vector $g_{ij}(k) = \{I_k\}$, $k = 1, \ldots, m$ where $I_k$ is an indicator variable that is equal to 1 if the glitch type $k$ is present, and 0 otherwise. Let $\omega_k$ be the user-supplied weights for the glitches.

The glitch index of a data set is simply the sum of the weighted glitch score of individual cells.

$$G(D) = \sum_{i,j,k} g_{ij}(k) \times \omega_k \quad (1)$$

The lower the glitch index, the "cleaner" the data set. For further details about glitch scoring, please see [3].

### 2.1.4 Statistical Distortion

While cleaning strategies remove glitches from data, they do not necessarily make the data more usable or useful, as illustrated in the example in Section 1. The data can be changed to such an extent that they no longer represent the underlying process that generated the data. To address this issue, we introduce the notion of *statistical distortion* which we define as follows:

*Definition 1.* If cleaning strategy $C$ applied to data set $\mathcal{D}$ yields a cleaned dataset $\mathcal{D}_C$, then the *statistical distortion* of $C$ on $\mathcal{D}$ is defined as:

$$\mathcal{S}(C, D) = d(\mathcal{D}, \mathcal{D}_C), \quad (2)$$

where $d(\mathcal{D}, \mathcal{D}_C)$ is a distance between the two empirical distributions. Possible distances are the Earth Mover's, Kullback-Liebler or Mahalanobis distances.

Ideally, we would like the improvement in the glitch index $G(D) - G(D_C)$ to be large, while keeping $\mathcal{S}(C, D) = d(\mathcal{D}, \mathcal{D}_C)$ small. In a perfect world, we could do both. For example, if a census has missing values, we can manually travel to all locations and collect the data. There would be no distortion, and we would have repaired the data. But it is not feasible, given the cost. Our experimental framework aims to help a user identify an acceptable trade-off, as discussed in reference to Figure 2.

### 2.1.5 Cost

The cost of implementing a data cleaning strategy is highly context dependent. In this paper, we use the percentage of glitches removed as a proxy for cost. The cost can be reflected more accurately by the user by assigning weights in a manner analogous to glitch weights. See [3] for further details.

### 2.1.6 Advantages of Our Framework

Our framework has several advantages: (1) Very large data: Our framework, because it relies on sampling, will work on very large data. (2) Customizable: It gives the end user flexibility in determining the ideal data set and glitch scoring mechanisms. Such flexibility is essential because of the highly domain and user dependent nature of data quality. (3) Cost control: The scale of the experiments can be determined by the resources available to the user. While a comprehensive and exhaustive experimental suite can identify an optimal cleaning strategy, even a limited set of experiments can provide insights into candidate strategies and their statistical impact on data. (4) Knowledge base: The experiments allow the user to understand the behavior of the strategies and match strategies to contexts and build a domain specific knowledge base for future use.

## 2.2 Paper Organization

The rest of the paper is organized as follows. In Section 3, we provide a detailed and specific formulation in the context of time series data collected on hierarchical data. We describe the experimental set up in Section 4. In Section 5, we discuss the findings from our experiments. We present our conclusions and opportunities for further research in Section 6.

## 3. FORMULATION

Our methodology is generally applicable to a wide spectrum of data ranging from structured data, hierarchical and spatio-temporal data, to unstructured data. The sampling mechanism can be designed to preserve the structure of the data. In this paper, we focus on a three-dimensional data stream accumulated on a hierarchical network.

## 3.1 Data Description

Let $N_i$, $N_{ij}$, $N_{ijk}$ represent nodes in successive layers of a network, i.e. node $N_{ijk}$ is part of node $N_{ij}$ which is in turn part of node $N_i$. Thus for example in the context of a mobility network, $N_i$ could represent the Radio Network Controller (RNC), $N_{ij}$, a cell tower (Node B) reporting to that RNC, and $N_{ijk}$ an individual antenna (sector) on that particular cell tower. The actual example is described in detail in [9]. However, for the purpose of this paper, it is sufficient to specify the structure of the data.

At each node, we measure $v$ variables in the form of a time series or data stream. For network node $N_{ijk}$, the data stream is represented by a $v \times 1$ vector, $\mathbf{X}_{ijk}^t$, where the subscript refers to the node, and the superscript to time. Let $\mathcal{F}_t$ denote the history up to time $t-1$. In the data stream context, it is often infeasible to store all the data. Analyses are restricted to the current window of the history, in conjunction with summaries stored from past history. In this paper we restrict ourselves to the currently available window, $\mathcal{F}_t^w$, the $w$ time-step history up to time $t-1$, for times $t-1, \ldots, t-w$.

## 3.2 Glitch Types

While data glitches come in a wide variety, our approach will work on any glitch that can be detected and flagged. The glitches can be multi-type, co-occurring or stand alone, with complex patterns of dependence as discussed in [3]. Our case study focuses on three types of glitches: missing and inconsistent values, and outliers.

## 3.3 Glitch Detection

We define a data glitch detector as a function on the data stream $\mathbf{X}^t$, where we ignore the subscript for clarity.

The *missing values* detector is given by:

$$f_M(\mathbf{X}^t) = \mathbf{I}_{\text{missing}},$$

where the vector $\mathbf{I}_{\text{missing}} \in (0, 1)^{v \times 1}$ and $\mathbf{I}_{\text{missing}}[i] = 1$ if $\mathbf{X}^t[i]$ is missing.

The glitch detectors for inconsistent values values are defined in a similar manner. An inconsistency can be defined based on a single attribute ("inconsistent if $X$ is less than 0"), or based on multiple attributes ("inconsistent if $X$ greater than 0 and $Y$ is less than 0"). Each inconsistency will result in a flag and contribute a bit to the inconsistent glitch vector $\mathbf{I}_{\text{inconsistent}}$. The methodology can handle multiple variants of a single glitch type, but for simplicity of notation, we will set a single flag for all inconsistency types, $\mathbf{I}_{\text{inconsistent}}[i] = 1$ or $0$. Therefore,

$$f_I(\mathbf{X}^t) = \mathbf{I}_{\text{inconsistent}},$$

where $\mathbf{I}_{\text{inconsistent}} \in (0, 1)^{v \times 1}$.

The outlier (anomaly) detection function $f_{\text{outlier}}$ typically is based on historical data prior to time $t$. It is of the form

$$f_O(\mathbf{X}^t | \mathbf{X}^{\mathcal{F}_t^w}, \mathbf{X}_N^{\mathcal{F}_t^w}),$$

where $\mathbf{X}_{\mathcal{F}_t^w}$ is the $w$ time-step history for the particular node, and $\mathbf{X}_N^{\mathcal{F}_t^w}$ is the $w$ time-step history for the neighbors of that node. The output of $f_O$ is a $v \times 1$ vector of 1's and 0's, that flags outliers for each of the $v$ attributes, based on a pre-selected threshold of outlyingness. Alternatively, the output of $f_O$ can be a vector of the actual $p$ values obtained from application of the outlier detection method. This gives the user flexibility to change the thresholds for outliers.

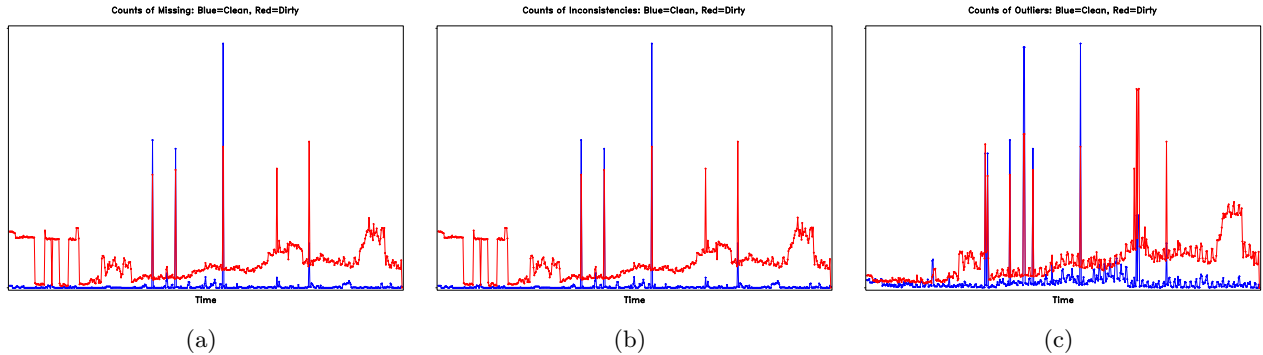

(a)  (b)  (c)

Figure 3: Time series of counts of three types of glitches aggregated across multiple runs (replications) and samples, roughly 5000 data points at any given time. There is considerable overlap between missing and inconsistent values. Experiments are conducted on time series of length 170 sampled from the data that generated the above glitch patterns.

Given the three types of glitches and the corresponding glitch detectors $f$, we construct for each node $N_{ijk}$ and for each time $t$, a $v \times 3$ bit matrix $G_{t,ijk}$, where

$$G_{t,ijk} = [f_{\mathrm{M}}(\mathbf{X}^t_{ijk}), f_{\mathrm{I}}(\mathbf{X}^t_{ijk}), f_{\mathrm{O}}(\mathbf{X}^t_{ijk}|\mathbf{X}^{\mathcal{F}^w_t}_{ijk}]\mathbf{X}^{\mathcal{F}^w_t}_{N_{ijk}})].$$

To summarize, given node $N_{ijk}$, the column $G_t[1,]$ is the bit vector that corresponds to all glitch types for variable 1 at time $t$; and the row $G_t[,2]$ corresponds to the occurrence of glitch type 2 at time $t$ across all $v$ variables.

### 3.4 Glitch Index

We define the overall glitch score of a dataset $\mathcal{D}$ as:

$$G(\mathcal{D}) = \left[I_{1 \times v}\left(\sum_{ijk}\sum_{t=1}^{T_{ijk}} G_{t,ijk}(\mathcal{D})/T_{ijk}\right)\right]W.$$

In the above expression $I_{1 \times v}$ is a $1 \times p$ vector of 1's, and $W$ is a $m \times 1$ weight vector that assigns user provided weights to each of the glitch types. In our example, $m = 3$. The data stream associated with any node $N_{ijk}$ may have different number of data points $T_{ijk}$, depending on whether the node was functional. The node level glitch scores $\sum_t G_{t,ijk}$ are normalized by $T_{ijk}$, to adjust for the amount of data available at each node, to ensure that it contributes equally to the overall glitch score.

### 3.5 Statistical Distortion: EMD

In Section 2.1.4 we defined *statistical distortion* as the distributional distance between two data sets. In this paper, we use Earth Mover's distance (EMD) to measure the distortion caused by our cleaning strategies. Earth Mover's distance, also known as Wasserstein's metric, and a special case of Mallows distance, [7] was popularized in computer vision and image retrieval community. Specifically, let $P$ and $Q$ be two distributions with the same support, and let $b_i, i = 1, \ldots n$ be the bins covering this support. Define

$$W(F; P, Q) = \sum_{i=1}^{n}\sum_{j=1}^{n} f_{ij}|b_i - b_j|,$$

where $F = \{f_{ij}\}$ represents the flow of density between bins $b_i$ and $b_j$ needed to convert distribution $P$ to $Q$.

If $F^* = \operatorname{argmin}_F W(F; P, Q)$, then the Earth Mover's distance is given by

$$\mathrm{EMD}(P, Q) = \sum_i \sum_j f^*_{ij}|b_i - b_j|/\sum_i \sum_j f^*_{ij}.$$

Thus EMD measures the difference between the distributions $P$ and $Q$, using cross-bin differences. Therefore, it is not affected by binning differences. It meaningfully matches the perceptual notion of "nearness". Recent work [1] and [13] have shown it to be computationally feasible.

## 4. EXPERIMENTS

We performed a series of experiments based on network monitoring data $\mathcal{D}$, to demonstrate our methodology. For the purpose of illustration, we use very simple strategies.

### 4.1 The Data Stream and Its Glitches

The network monitoring data consist of 20,000 time series, each of length at most 170, measured on three variables. Each stream corresponds to an antenna (sector) on a cell tower. We annotated each data point with bit vectors that indicate the presence or absence of missing values, inconsistencies and outliers. A value is considered missing if it is not populated. Inconsistencies are specified as constraints: "(1) Attribute 1 should be greater than or equal to zero, (2) Attribute 3 should lie in the interval $[0, 1]$, and (3) Attribute 1 should not be populated if Attribute 3 is missing." A violation of any of these three constraints is an inconsistency. Outliers are identified using 3-$\sigma$ limits on an attribute by attribute basis, where the limits are computed using ideal data set $\mathcal{D}_I$. The latter was composed of data from sectors where the time series contained less than 5% each of missing, inconsistencies and outliers.

### 4.2 Sampling and Simulation

We sampled with replacement from $\mathcal{D}$ and $\mathcal{D}_I$ to simulate the test pairs $\mathcal{D}^i, \mathcal{D}^i_I$. We created 50 such pairs. We tried different sample sizes for the data sets in the pair, but beyond a certain point, the sample size had no significant impact on our results. The results shown in the rest of the paper are for sample size=100 and 500 pairs of time series, of length 170, and with three attributes (a 3-tuple). There were $2 \times 50 \times 100 \times 170 = 1,700,000$ individual 3-tuples for the experiments with sample size 100; and $2 \times 50 \times 500 \times 170 = 8,500,000$ 3-tuples for the experiments conducted with sample size 500. We maintained the temporal structure by sampling entire time series and not individual data points. This allows us to study temporal correlations among glitches. In future, we plan to use a sampling scheme that preserves not just temporal structure, but also network topology hierarchy.

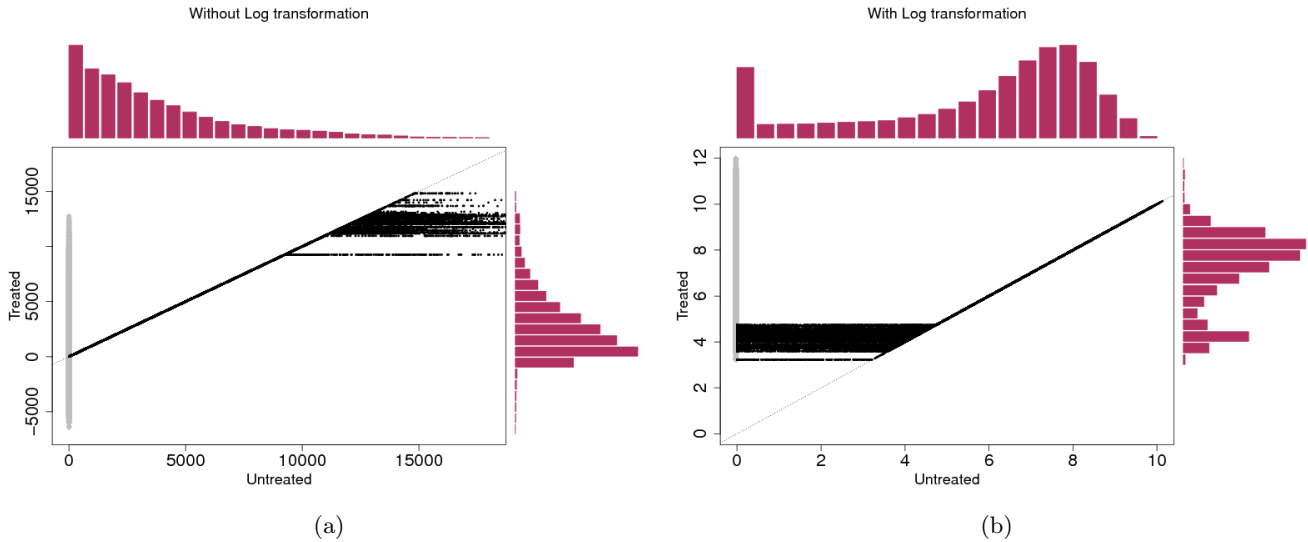

Figure 4: Scatter plots of Attribute 1, untreated vs. treated by applying Strategy 1. The histograms depict distributions of Attribute 1 (a) without any transformation, and (b) with a log transformation before cleaning. The gray dots near the $Y$-axis represent missing values imputed by SAS/PROC MI. Black dots along the $y=x$ line are the data untouched by the cleaning strategy. The horizontal black dots correspond to winsorized values. Each horizontal line corresponds to a separate run, and the variability is caused by sampling variation of the 3-$\sigma$ limits used for winsorization. Note that the imputing algorithm creates new inconsistencies during imputation by allowing the treated values to be negative in (a), because it assumes an underlying Gaussian distribution that is not appropriate for this data.

Figure 3 shows time series depicting the counts of each of the three types of glitches, in 50 runs of 100 samples each. That is, at any given point in time, the glitches are aggregated over 5000 series. The three plots are on the same scale and are comparable. There is considerable overlap between missing and inconsistencies, indicating that they are co-occurring glitches.

*Factors:* We conducted experiments to study the effects of: (1) attribute transformations, (2) different strategies, (3) cost, and (4) sample size.

## 5. ANALYSIS

In the following section we describe the specifics of our experiments, their outcome, and interpretation of results.

### 5.1 Cleaning Strategies

For the purpose of illustration, we applied five cleaning strategies to each of the 100 test pairs (samples) of data streams. We repeated this 50 times. Strategy 1 imputes values to missing and inconsistent data using SAS PROC MI, and outliers are repaired by winsorization on an attribute by attribute basis. Strategy 2 ignores outliers, but treats missing and inconsistent values using SAS PROC MI. Strategy 3 ignores missing and inconsistent values, but treats outliers using winsorization. Strategy 4 ignores outliers, but treats missing and inconsistent values by replacing them with the mean of the attribute computed from the ideal data set. Strategy 5 treats all three types of glitches by replacing missing and inconsistent with mean values from the ideal data set, and by winsorizing the outliers. We also study the effect of transformations on cleaning strategies by applying a natural log transformation to one of the variables.

We computed glitch improvement $G(\mathcal{D}^i) - G(\mathcal{D}^i_I)$, with a weight of 0.25 each to missing and inconsistent values, and 0.5 to outlier glitches. We calculated statistical distortion $S(\mathcal{D}^i_C, \mathcal{D}^i)$ using the EMD between dirty and treated data, as discussed in Section 3.5.

### 5.2 Studying Cost

Since the cost of a cleaning strategy depends on the user's resources, we used the proportion of glitches cleaned as a proxy for cost. For each time series, we computed the normalized glitch score, and ranked all the series in the dirty data set by glitch score. We applied the cleaning strategy to the top $x\%$ of the time series, and measured the improvement. When $x = 0$, the treated data is identical to the dirty data, and when $x = 100$, all the glitches are cleaned (note that new glitches may be introduced in the process). Ideally, beyond a certain threshold of $x$, incremental cleaning should yield no additional advantages because statistical distortion may increase rapidly.

### 5.3 Data Transformation and Cleaning

Figures 4 shows Attribute 1, where we plot untreated values against those treated by applying cleaning Strategy 1,

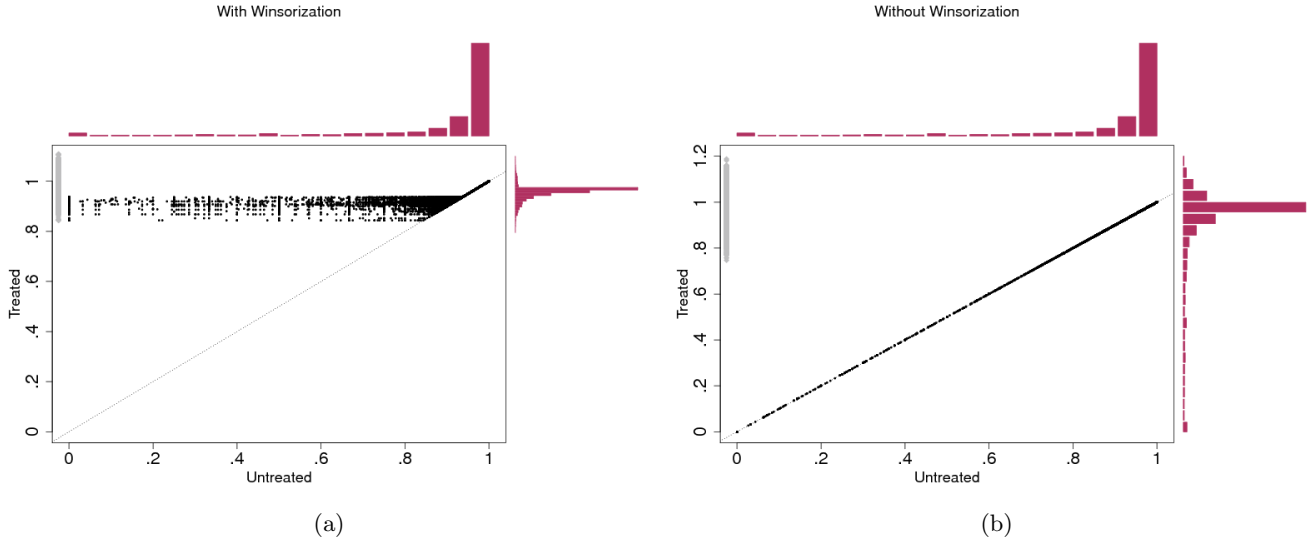

Figure 5: Scatter plots of Attribute 3 before and after applying strategies 1 and 2 respectively. Imputed values fall within a range close to 1. Values greater than 1, which do not occur in the original data set, are imputed by the imputing algorithm, resulting in new inconsistencies.

(a) without and (b) with the log transformation. The histograms at the axes show the corresponding data distribution. The plots are generated using the data in all 50 replications. The gray points represent instances where the variable was missing in the original data set. The imputation algorithm imputes values according to some assumed distribution, typically a multivariate Gaussian. The black points lying along the $y = x$ line correspond to instances where the values are untouched. The black points outside $y = x$ are Winsorized values, imputed with the most extreme value acceptable. There is considerable sampling variability in the Winsorized values between experimental runs. Without the log transformation, some of the imputed values are negative, because the imputing algorithm assumes a Gaussian, resulting in inconsistencies. This is an instance where a cleaning strategy introducing new glitches.

Winsorization effects the two cases in a different manner. Without the log transformation, the distribution is skewed to the right. Large values are identified as outliers and Winsorized. With the log transformation, the distribution is skewed to the left and the lower values are Winsorized instead. This a cautionary tale against the blind use of attribute transformations and quantitative cleaning strategies.

### 5.4 Strategies and Attribute Distributions

Figures 5 (a)-(b) show similar plots for Attribute 3, for cleaning Strategy 1 and 2 respectively. In both cases, the imputed values fall within a range close to 1, where the bulk of the original values lie. Values greater than 1, which do not occur in the original data set, are imputed by the imputing algorithm, resulting in new inconsistencies. In Figure 5a, there is variability in the Winsorized values, but all fall in the range of 0.85 to 0.95. Strategy 2 ignores outliers so the new inconsistent values remain untreated, and the algorithm imputes values over the entire range, all the way to 0 (Figure 5b).

### 5.5 Evaluating Strategies

Figure 6 plots statistical distortion (EMD) against the improvement in glitch scores, $G(\mathcal{D}) - G(\mathcal{D}_C)$, for treated data set $\mathcal{D}_C$, and the original data set $\mathcal{D}$. Figure 6(a) is based on 50 replications, each consisting of 100 pairs of time series, with log transformation applied to Attribute 1; Figure 6(b) without the log transformation; and Figure 6(c) with the log transformation but with a sample size of 500 time series in each replication.

We applied each of the five cleaning strategies discussed in Section 5.1, to each of the three cases. The effects of these strategies on statistical distortion and glitch score improvement are shown, where each point corresponds to a test pair $(\mathcal{D}^i, \mathcal{D}_C{}^i)$.

Based on these two performance metrics, we find that there is a clear separation of the strategies. First, consider the use of a single cleaning method for both missing and inconsistent values, through imputation by (1) SAS PROC MI (black dot) or (2) replacement with the mean (blue circle). Both resulted in similar glitch improvement because they treat the same glitches. However, replacement with the mean value resulted in a lower statistical distortion. This is due to the fact that the imputation algorithm assumes a symmetric distribution that is not appropriate for the highly skewed attribute distributions at hand.

Next, consider treating just the outliers by Winsorization (green triangles). The log transformation of Attribute 1 results in a greater number of outliers based on the 3-$\sigma$ threshold. Therefore, by treating more glitches, we achieved a greater glitch score improvement when the log transformation was applied, as seen in Figure 6 (a) and (c).

Next, consider simultaneously using two cleaning methods, one for missing and inconsistent values, and one for outliers. The result is an expected improvement in glitch scores compared to using a single cleaning method. Here too, replacing missing and inconsistent values with mean

1680

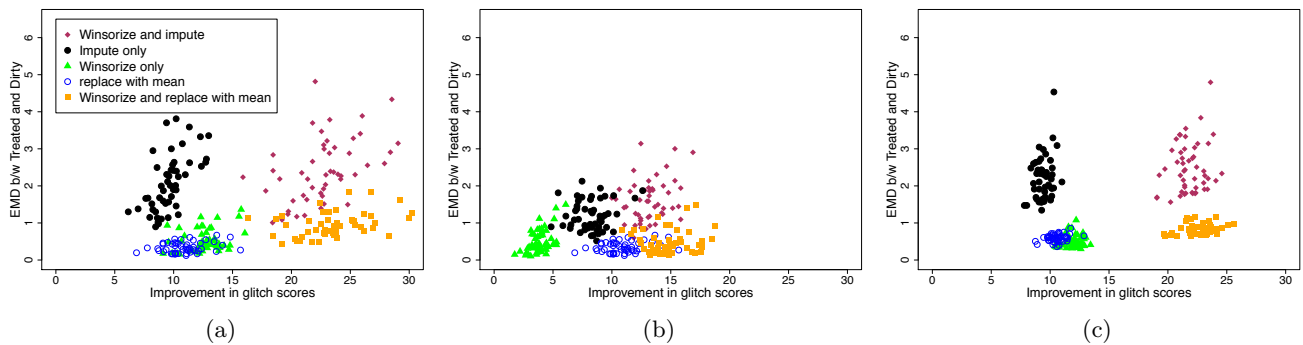

Figure 6: Scatter plots of statistical distortion and glitch score improvement for five cleaning strategies applied to each of 50 experimental runs. (a) sample size of 100 pairs of time series with the log transformation applied to Attribute 1; (b) sample size of 100 pairs of time series without the log transformation; (c) sample size of 500 pairs of time series with log transformation.

(orange squares) causes less distortion than imputing using the PROC MI algorithm (red diamonds).

We find that in all three cases Figure 6 (a)-(c), the relative performance of the strategies is roughly the same. The exception is the strategy represented by the green triangles, where we treat just the outliers using Winsorization. The ordering is changed by the effect of log transformation that results in a different tail of the distribution being Winsorized, as discussed earlier.

Lastly, we find that with an increase in sample size, the points coalesce, separating the strategies more crisply.

## 5.6 Cost of Cleaning

We investigated the effect of the cost of a cleaning strategy on the resulting glitch score and statistical distortion. Instead of directly accounting for the computational cost, we measured the cost by the amount of data that is cleaned. Specifically, we ranked each time series according to its aggregated and normalized glitch score, and cleaned the data from the highest glitch score, until a pre-determined proportion of the data was cleaned. We used the strategy of Winsorization of outliers and imputation of missing/inconsistent values. Figure 7 shows the results.

The red-colored diamonds correspond to case when all the data is cleaned. Other points correspond to the cleaning strategy applied to 50% (blue triangles), 20% (green squares) and 0% (yellow dots) of the data. As expected, less data we clean, the less the improvement in glitch score, However, the amount of statistical distortion is also less. In the extreme case of 0% data cleaned, there is no change in the glitch score, and no statistical distortion relative to the original data. (Note: In Figure 7(b), glitch score improvement for untreated data is not identically 0. This is an artifact because when there is a tie for the glitch score cutoff, the sampling scheme retrieves the data from the treated data rather than the dirty data.)

In our example, Figure 7 suggests that cleaning more than 50% of the data results in relatively small changes in statistical distortion and glitch score. If cost is important, then it might be enough to clean only 50% of the data. However, if it is essential to clean the entire data due to a corporate or legal mandate, the strategy that results in maximum glitch score improvement with the least distortion and cost should be chosen.

Table 5.6 summarizes our findings. There is a strong overlap between missing values and inconsistencies (some due to our very definition of the inconsistency constraint). Increasing the sample size results in a tightening of the values.

In our example, a simple and cheap strategy outperformed a more sophisticated and expensive strategy mainly because the latter made assumptions that were not suitable for the data. Our aim in in conducting these experiments was to illustrate our methodology and experimental framework, rather than make any specific recommendations for a particular application.

## 6. CONCLUSION

In this paper, we proposed a three-dimensional approach to evaluating data cleaning strategies. While existing approaches focus mostly on glitch improvement and cost, we introduced the additional important notion of statistical distortion. Data cleaning alters the underlying data distributions; when these changes distort the data to such an extent that they no longer truly represent the underlying process that generated the data, data cleaning has gone too far, rendering the data unusable. We used the Earth Mover Distance (EMD) to measure statistical distortion. We proposed an experimental framework based on sampling to evaluate cleaning strategies using the three criteria of glitch improvement, statistical distortion and cost.

To our surprise, a simple imputation method outperformed a more sophisticated algorithm that relied on assumptions that were not suitable for the data.

## 6.1 Spatio-Temporal Glitch Correlations

We demonstrated our framework using data streams where we preserved the temporal structure by sampling entire time series. However, we did not preserve network topology. This is important since glitches tend to cluster both temporally as well as topologically (spatially) because they are often driven by physical phenomena related to collocated equipment like antennae on a cell tower. Our future work focuses on developing sampling schemes for preserving network topology, and for studying cleaning strategies that address spatio-temporal correlations. Cleaning algorithms that make use of the correlated data cost less and perform better, resulting in less statistical distortion. We will also develop special glitch scoring and statistical distortion measures to address spatio-temporal glitch correlations. One approach consists of treating the glitch sequences as a multivariate counting process, and computing glitch score index based on the realizations of the counting process.

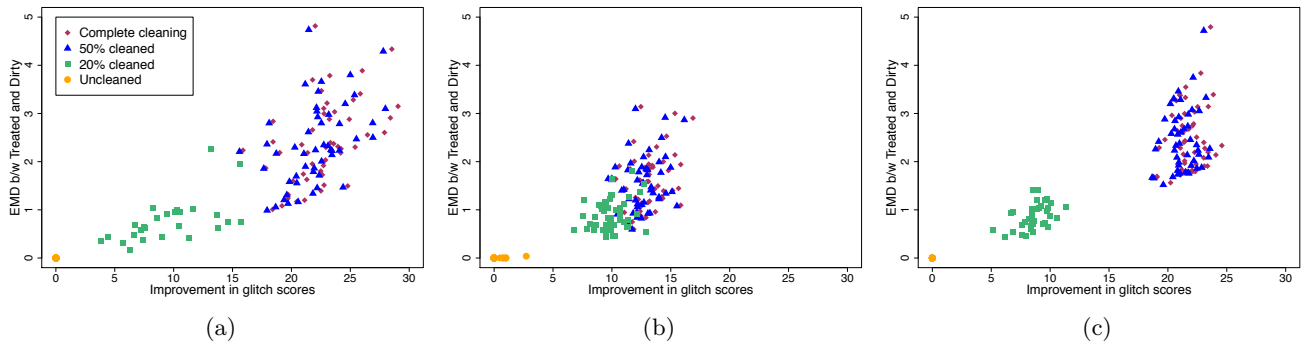

Figure 7: Scatter plots of statistical distortion and glitch score improvement for each of 50 experimental replications. Strategy 1 is used for cleaning 100%, 50%, 20% and 0% of the data: (a) Sample size of 100 pairs of time series with the log transformation applied to Attribute 1. (b) Sample size of 100 without the log transformation. (c) Sample size of 500 pairs of time series with the log transformation.

Table 1: Percentage of Glitches in 50 Runs:Before and After Cleaning

|  | Cleaning Strategy | Missing Dirty | Inconsistent Dirty | Outliers Dirty | Missing Treated | Inconsistent Treated | Outliers Treated |
|---|---|---|---|---|---|---|---|
| n=100, log(attribute 1) | Strategy 1 | 15.7965 | 15.8822 | 16.7723 | 0.0281166 | 2.19791 | 0 |
|  | Strategy 2 | 15.7965 | 15.8822 | 16.7723 | 0.0281166 | 2.19791 | 17.6094 |
|  | Strategy 3 | 15.7965 | 15.8822 | 16.7723 | 15.7965 | 15.8822 | 0 |
|  | Strategy 4 | 15.7965 | 15.8822 | 16.7723 | 0 | 0 | 16.7723 |
|  | Strategy 5 | 15.7965 | 15.8822 | 16.7723 | 0 | 0 | 0 |
| n=500, log(attribute 1) | Strategy 1 | 15.2175 | 15.3082 | 15.8594 | 0.0262495 | 1.97544 | 0 |
|  | Strategy 2 | 15.2175 | 15.3082 | 15.8594 | 0.0262495 | 1.97544 | 16.611 |
|  | Strategy 3 | 15.2175 | 15.3082 | 15.8594 | 15.2175 | 15.3082 | 0 |
|  | Strategy 4 | 15.2175 | 15.3082 | 15.8594 | 0 | 0 | 15.8594 |
|  | Strategy 5 | 15.2175 | 15.3082 | 15.8594 | 0 | 0 | 0 |
| n=100, no log | Strategy 1 | 15.7965 | 15.8822 | 5.07022 | 0.0281166 | 4.75612 | 0 |
|  | Strategy 2 | 15.7965 | 15.8822 | 5.07022 | 0.0281166 | 4.75612 | 6.40656 |
|  | Strategy 3 | 15.7965 | 15.8822 | 5.07022 | 15.7965 | 15.8822 | 0 |
|  | Strategy 4 | 15.7965 | 15.8822 | 5.07022 | 0 | 0 | 5.07022 |
|  | Strategy 5 | 15.7965 | 15.8822 | 5.07022 | 0 | 0 | 0 |

In addition, we will investigate other ways of computing the distance. In this paper, while we sampled entire time series, we computed EMD treating each time instance as a separate data point. We will investigate computationally feasible distance measures suitable for spatio-temporal glitch cleaning.

Finally, we plan to conduct additional experiments and test our statistical framework on additional data. We will also investigate the performance of data repair techniques proposed in existing literature, using our three dimensional data quality metric that includes statistical distortion, glitch improvement and cost.

## 7. REFERENCES

[1] D. Applegate, T. Dasu, S. Krishnan, and S. Urbanek. Unsupervised clustering of multidimensional distributions using earth mover distance. In *KDD*, pages 636–644, 2011.

[2] C. Batini, C. Cappiello, C. Francalanci, and A. Maurino. Methodologies for data quality assessment and improvement. *ACM Comput. Surv.*, 41(3):16:1–16:52, 2009.

[3] L. Berti-Equille, T. Dasu, and D. Srivastava. Discovery of complex glitch patterns: A novel approach to quantitative data cleaning. In *ICDE*, pages 733–744, 2011.

[4] E. Cohen and H. Kaplan. Summarizing data using bottom-k sketches. In *PODC*, pages 225–234, 2007.

[5] N. Duffield, C. Lund, and M. Thorup. Priority sampling for estimation of arbitrary subset sums. *J. ACM*, 54(6):32:1–32:37, 2007.

[6] W. Fan, F. Geerts, X. Jia, and A. Kementsietsidis. Conditional functional dependencies for capturing

data inconsistencies. *ACM Trans. Database Syst.*, 33(2):6:1–6:48, 2008.
[7] E. Levina and P. Bickel. The earth movers distance is the mallows distance: some insights from statistics. In *ICCV*, pages 251–256, 2001.
[8] X. Liu, X. Dong, B. C. Ooi, and D. Srivastava. Online data fusion. In *VLDB*, pages 932–943, 2011.
[9] J. Loh and T. Dasu. Auditing data streams for correlated glitches. In *ICIQ*, pages 1–15, 2011.
[10] C. Mayfield, J. Neville, and S. Prabhakar. Eracer: a database approach for statistical inference and data cleaning. In *SIGMOD*, pages 75–86, 2010.
[11] F. Olken. Random sampling from databases. *PhD. thesis, University of California, Berkeley*, 1993.
[12] L. L. Pipino, Y. W. Lee, and R. Y. Wang. Data quality assessment. *Commun. ACM*, 45(4):211–218, 2002.
[13] S. Shirdhonkar and D. W. Jacobs. Approximate earth mover's distance in linear time. In *CVPR*, pages 1–8, 2008.
[14] M. Yakout, A. K. Elmagarmid, J. Neville, and M. Ouzzani. Gdr: a system for guided data repair. In *SIGMOD*, pages 1223–1226, 2010.